\documentstyle[aps,psfig,graphics,multicol]{revtex}
\begin{document}
\draft

\title{
Self-avoiding walks on scale-free networks}
\author{Carlos P. Herrero}
\address{
    Instituto de Ciencia de Materiales,
    Consejo Superior de Investigaciones Cient\'{\i}ficas (CSIC),
    Campus de Cantoblanco, 28049 Madrid, Spain \\   }
\date{\today}
\maketitle

\begin{abstract}
Several kinds of walks on complex networks are currently used to
analyze search and navigation in different systems. Many analytical and 
computational results are known for random walks on such networks.
Self-avoiding walks (SAWs) are expected to be more suitable than unrestricted
random walks to explore various kinds of real-life networks.
Here we study long-range properties of random SAWs
on scale-free networks, characterized by a degree distribution 
$P(k) \sim k^{-\gamma}$.
In the limit of large networks (system size $N \to \infty$), the average 
number $s_n$ of SAWs starting from a generic site increases as
$\mu^n$, with $\mu = \langle k^2 \rangle / \langle k \rangle - 1$.
For finite $N$, $s_n$ is reduced due to the presence of loops in the network,
which causes the emergence of attrition of the paths. For kinetic growth walks,
the average maximum length, $\langle L \rangle$, increases as 
a power of the system size: $\langle L \rangle \sim N^{\alpha}$, 
with an exponent $\alpha$ increasing as the parameter $\gamma$ is raised. 
We discuss the dependence of $\alpha$ on the minimum allowed degree in 
the network.  A similar power-law dependence is found for the mean 
self-intersection length of non-reversal random walks.
Simulation results support our approximate analytical calculations.
\end{abstract}

\pacs{PACS numbers: 89.75.Fb, 87.23.Ge, 05.40.Fb, 89.75.Da}
%
% 05.40.Fb Random walks and Levy flights
% 87.23.Ge Dynamics of social systems 
% 05.50.+q Lattice theory and statistics (Ising, Potts, etc.)
% 07.05.Mh Neural networks, fuzzy logic, artificial intelligence
% 84.35.+i Neural networks (Electronics) 
% 87.18.Sn Neural networks (Biological and medical physics)
% 89.75.Da Systems obeying scaling laws
% 89.75.Fb Structures and organization in complex systems
% 89.75.Hc Networks and genealogical trees

\begin{multicols}{2}
                                                                                    
\section{Introduction}
Many natural and artificial systems have a network structure,
where nodes represent typical system units and edges represent interactions 
between connected pairs of units. 
Thus, complex networks are currently used to model several kinds of real-life 
systems (social, biological, technological, economic), and to study different 
processes taking place on them \cite{st01,al02,do02a}.
In recent years, new models of complex networks have been designed to explain
empirical data in several fields.
This is the case of the so-called small-world \cite{wa98} and scale-free
networks \cite{ba99}, which  incorporate various aspects of real systems.
These complex networks provide us with the underlying topological structure to
analyze processes such as spread of infections \cite{mo00,ku01}, signal propagation
\cite{wa98,he02}, and random spreading of information \cite{pa01,la01b}.
They have been also employed to study statistical physical problems
as percolation \cite{mo00,ne99} and cooperative phenomena \cite{ba00,le02}.

In a scale-free (SF) network the degree distribution $P(k)$, where $k$ is the 
number of links connected to a node, has a power-law decay $P(k) \sim k^{-\gamma}$.
This kind of networks have been found in social systems \cite{ne01},
for protein interactions \cite{je01}, in the internet \cite{si03},
and in the world-wide web \cite{al99}.
In both natural and artificial networks, the exponent $\gamma$ controlling the
degree distribution is usually in the range $2 < \gamma < 3$ \cite{do02a,go02}.
The origin of such power-law degree distributions was addressed by
Barab\'asi and Albert \cite{ba99}, who argued that two ingredients are
sufficient to explain the scale-free character of many real-life networks,
namely: growth and preferential attachment. They found that the combination of 
both criteria yields non-equilibrium SF networks with an exponent $\gamma = 3$.
One can also study equilibrium SF networks, defined as statistical ensembles
of random networks with a given degree distribution 
$P(k) \sim k^{-\gamma} $\cite{do02a}, for which one can analyze several
properties as a function of the exponent $\gamma$.
SF networks display the so-called small-world effect, and they have
been found to be ultrasmall, in the sense that the mean distance between sites
increases with the network size $N$ slower than $\log N$ \cite{co03}.

 Social networks form the substrate where dynamical processes such as disease 
propagation and information spreading take place. These networks have the property 
of being searchable, i.e. people (nodes in a network) can direct messages through 
their network of acquaintances to reach a distant specific target in only 
a few steps \cite{wa02,gu02,ki02}.
It is clear that the structure of such networks will play an important role 
in these dynamical processes, which are usually studied by means of stochastic 
dynamics and random walks. 
Several characteristics of random walks on complex networks have been analyzed 
in connection with diffusion and exploration processes \cite{je00,ta01,ne03}.
In this context, it is known that some processes, such as navigation and 
exploratory behavior are neither purely random nor totally deterministic
\cite{li01}, and can be described by walks on graphs \cite{sa01,ad01}.  

Self-avoiding walks (SAWs) can be more effective than unrestricted random walks in 
exploring a network, since they cannot return to sites already visited. This 
property has been used by Adamic {\em et al.} \cite{ad01} to define local search 
strategies in scale-free networks.
However, the self-avoiding property causes attrition of the paths, in the
sense that a large fraction of paths generated in a stochastic manner have
to be abandoned because they are overlapping. This can be a serious limitation 
to explore networks with pure SAWs.

SAWs have been traditionally used to model structural and dynamical properties 
of macromolecules \cite{ge79,le89}. They are also useful to characterize complex 
crystal structures \cite{he95} and to study critical phenomena in lattice 
models \cite{kr81}.  Universal constants for SAWs on regular lattices have been 
discussed by Privman {\em et al.} \cite{pr91}. In our context of complex networks,
the asymptotic properties of SAWs have been studied recently in small-world 
networks \cite{he03}.

Here we study long-range properties of SAWs on equilibrium scale-free networks,
and discuss the `attrition problem'.  The number of surviving walks to a given 
length $n$ is obtained by an approximate analytical procedure, and the results 
are compared with those obtained from numerical simulations. 
In particular, we find that the number of surviving walks after $n$ steps 
scales as a power of the system size $N$. 
We note that the term `length' is employed throughout this paper to
indicate the (dimensionless) number of steps of a walk, as usually in the
literature on networks \cite{do02a}.

The paper is organized as follows. 
In Sec.\,II we give some definitions and concepts related to SAWs, along
with details on our computational method.
In Sec.\,III we calculate the number and end-to-end separation of SAWs in 
(uncorrelated) scale-free networks.  In Sec.\,IV we analyze the length at 
which non-reversal random walks intersect themselves in these networks 
(self-intersection length), and in Sec.\,V we calculate the average attrition 
length of kinetic growth SAWs, at which
they cannot continue without violating the self-avoidance condition.
The paper closes with some conclusions in Sec.\,VI.

\section{Basic definitions and method}
A self-avoiding walk (SAW) is defined as a walk along the bonds of
a given network which can never intersect itself. The walk is
restricted to moving to a nearest-neighbor site during each step,
and the self-avoiding condition constrains the walk to occupy only
sites which have not been previously visited in the same walk.

The simplest procedure to obtain SAWs consists just in generating ordinary
random walks and stop when they arrive at a node already visited.
A problem with this sampling algorithm in regular lattices is the
exponentially rapid attrition for long walks, since the probability of an
$n$-step walk being self-avoiding behaves for large $n$ as
$e^{- \lambda n}$, where $\lambda$ is the so-called attrition constant
\cite{so95}.
Due to this limitation, more sophisticated schemes based on Monte Carlo sampling, 
have been employed to generate SAWs with the correct weight, and to obtain ensemble 
averages of several quantities \cite{so95}. This has allowed, for example, 
to model the equilibrium statistics of linear polymers in dilute solutions.
 In general, for networks including nodes with different degrees (contrary to 
usual regular lattices), sampling by using simple random walks introduces 
a bias in the weight of different SAWs.

\begin{figure}
%\vspace*{-1.7cm}
\vspace*{-4.7cm}
\centerline{\psfig{figure=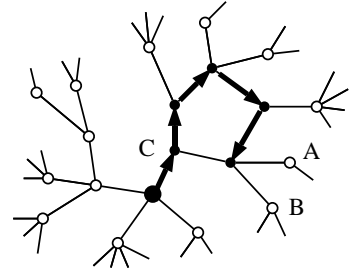,height=12.0cm}}
\vspace*{-3.4cm}
\caption{
Schematic diagram showing a non-reversal random walk of length $n = 5$ on a
realization of a random graph. Open and black circles represent unvisited
and visited nodes, respectively. The starting node is indicated by a larger
circle. The non-reversal condition allows in principle for the next (sixth)
step three possible nodes (denoted $A$, $B$, and $C$). The self-avoiding
condition excludes node $C$ for the sixth step. For a non-reversal SAW one
chooses among nodes $A$, $B$, and $C$. If $C$ is selected, then the walk stops.
For a kinetic growth walk, one chooses $A$ or $B$.
} \label{f1} \end{figure}

One can also consider kinetically grown SAWs, which can be more adequate to
analyze dynamic processes. Such walks are well-suited to study, for example, 
search or navigation processes on networks, where they are assumed to
grow step by step in a temporal sequence.
In the following we will consider two kinds of growing walks. The first kind 
will be `non-reversal' self-avoiding walks \cite{so95}. In these walks one randomly 
chooses the next step from among the neighboring nodes, excluding the previous 
one. If it happens that one chooses an already visited node, then the walk 
stops (see Fig. 1). These walks will allow us to study the `self-intersection 
length' (see Sec. IV).  
The second kind of walks considered here are kinetic growth walks
\cite{ma84}, in which one randomly chooses the next step among the
neighboring unvisited sites and stops growing when none are available. 
These walks were studied to describe the irreversible growth of linear
polymers \cite{ma84}, and will allow us to consider the `attrition length' 
for a walk on a given network (see Sec. V).   
Note that kinetic growth walks are less sensitive to attrition than
non-reversal SAWs, since in the former the walker always escapes whenever 
a way exists. 

We consider SF networks with degree distribution $P(k) \sim k^{-\gamma}$.
They are characterized, apart from the exponent $\gamma$ and the system size $N$, 
by the minimum degree $k_0$, which affects markedly some characteristics of SAWs 
in these networks (see below). We assume that $P(k) = 0$ for $k < k_0$.
Our networks are uncorrelated, in the sense that degrees of
nearest neighbors are statistically independent.
This means that the joint probability $P(k,k')$ fulfills the relation \cite{do02a}
\begin{equation}
  P(k,k') = \frac{k \, k'}{\langle k \rangle^2}  P(k) P(k') \, .
\label{kk1}
\end{equation}

For the numerical simulations we have generated networks with several
values of $\gamma$, $k_0$, and $N$. 
To generate a network, once defined the number of nodes $N_k$ with degree
$k$, we ascribe a degree to each node according to the set $\{N_k\}$, and 
then connect at random ends of links (giving a total of $L = \sum_k k N_k/2$ 
connections), with the conditions:
(i) no two nodes can have more than one bond connecting them, and
(ii) no node can be connected by a link to itself.
We have checked that networks generated in this way are uncorrelated, i.e.
they fulfill Eq. (\ref{kk1}).
All networks considered here contain a single component, i.e. any node in a 
network can be reached from any other node by traveling through a finite number 
of links. 
For each set of parameters ($\gamma$, $k_0$, $N$), we considered different
network realizations, and for a given network we selected at random the
starting nodes for the SAWs.  
For each considered parameter set, the total number of generated SAWs 
amounted to about $5 \times 10^5$.

 For regular lattices, the number $s_n$ of different SAWs starting
from a generic site has an asymptotic dependence for large $n$
\cite{pr91}: $s_n \sim n^{\Gamma - 1} \mu^n$,
where $\Gamma$ is a critical exponent which depends on the lattice dimension,
and $\mu$ is the so-called `connective constant' or effective coordination
number of the considered lattice \cite{ra85}.
In general, for a lattice with connectivity $k_0$, one has $\mu \le k_0 - 1$.
This parameter $\mu$ can be obtained as the limit
\begin{equation}
       \mu = \lim_{n\to\infty} \frac{s_n}{s_{n-1}}  \hspace{3mm} .
\label{mulim}
\end{equation}            
The connective constant depends upon the particular topology of each 
lattice, and has been determined very accurately for two- and three-dimensional 
lattices \cite{so95}. 

 For poissonian and scale-free networks the number of SAWs of length $n$
depends on the considered starting node of the network.
In the sequel we will call $s_n$ the average number of SAWs of length $n$,
i.e. the mean value obtained (for each $n$) by averaging over the 
network sites and over different network realizations (for given $\gamma$, 
$k_0$, and $N$).
For Erd\"os-R\'enyi random networks with poissonian distribution of degrees,
one has $s_n^{rd} = \langle k \rangle^n$ \cite{he03}, and therefore the connective 
constant is $\mu = \langle k \rangle$.
In connection with this, we note that for a Bethe lattice (or Cayley
tree) with connectivity $k_0$, the number of SAWs is given
by  $s_n^{BL} = k_0 (k_0-1)^{n-1}$, and one has $\mu_{BL} = k_0 - 1$. 

\section{General characteristics of SAWs}
\subsection{Number of walks}
 We calculate first the average number, $r_n$, of different $n$-step 
unrestricted walks starting from a node chosen at random. 
One trivially has $r_1 = \langle k \rangle$.
To calculate $r_n$ for $n>1$, one needs the degree distribution for nodes at 
which one arrives following a random edge.
Thus, given a generic node and a link starting on it, we call $Q(k)$
the degree distribution for the other end of the link.
The probability of reaching a node with connectivity $k$ is proportional
to $k$; therefore
\begin{equation}
Q(k) = \frac{k}{\langle k \rangle} \; P(k) \;  ,
\label{qk}
\end{equation}
where $\langle k \rangle$ in the denominator is a normalization factor.
Then, the average number of two-step random walks is given by
$r_2 = r_1 \langle k \rangle_Q$, where the subscript $Q$ indicates that the 
average value is taken with the probability distribution $Q(k)$. We find
$r_2 = \langle k^2 \rangle$ [average values without subscripts are taken
with the degree distribution $P(k)$].
For $n>2$ we have $r_n = r_{n-1} \langle k \rangle_Q$, and then
\begin{equation}
r_n = \langle k \rangle  \left( \frac{\langle k^2 \rangle}{\langle k \rangle}
          \right)^{n-1}   \; .
\label{rn}
\end{equation}
We have checked that this expression for $r_n$ and those given below for SAWs 
coincide with those derived by using a generating function for the degree 
distribution \cite{ad01,ca00}.

We now calculate the average number, $s_n$, of different self-avoiding walks of $n$ 
steps starting from a node taken at random, in uncorrelated networks. We will 
first consider the case $n/N \to 0$ (thermodynamic limit). For $n=1$, one has 
$s_1 =\langle k \rangle$.
For $n > 1$ we take into account that each ($n-1$)-step walk arriving at a node
with degree $k$, gives rise to $k-1$ $n$-step walks. Thus, we have
$s_n = s_{n-1} \langle k-1 \rangle_Q$, which yields
\begin{equation}
s_n = \langle k \rangle  \left( \frac{\langle k^2 \rangle}{\langle k \rangle} - 1
          \right)^{n-1}   \; . 
\label{sn}
\end{equation}
Then, the connective constant $\mu_{\infty}$ for $N \to \infty$ is given by
\begin{equation}
\mu_{\infty} = \frac{\langle k^2 \rangle}{\langle k \rangle} - 1   \;  .
\label{mu}
\end{equation}
This is consistent with the fact that uncorrelated networks are locally tree-like,
and the ratio $\langle k^2 \rangle / \langle k \rangle$ is the average degree
of a randomly-chosen end node of a randomly chosen link \cite{do02a}. 
For SF networks with $\gamma \leq 3$, $\mu$  diverges
as $N \to \infty$, as a consequence of the divergence of $\langle k^2 \rangle$.
For $\gamma > 3$, we can approximate the average values in Eq. (\ref{mu})
by replacing sums by integrals, and find
\begin{equation}
\mu_{\infty} \approx k_0 \; \frac{\gamma - 2}{\gamma - 3} - 1   \;  .
\label{mu2}
\end{equation}

Note that the ratio $s_n / s_{n-1}$ does not depend on $n$ for system size 
$N \to \infty$.  This is equivalent to assume that nodes in different steps of a 
non-reversal random walk are different. 
This means, in other words, that the probability of finding loops with 
$n'\le n$ in a $n$-step walk is negligible.
For finite networks, however, there will appear loops of any size \cite{bi03}, 
introducing corrections to the number of SAWs, and $s_n$ will be lower than 
given by Eq. \ref{sn}. 
These corrections will be of order $n/N$ for $n/N \ll 1$. 
The effects of this reduction in the number of random SAWs in scale-free 
networks will be considered in Sections IV and V.

 As indicated above, the number of SAWs on regular lattices scales for large 
$n$ as $s_n \sim n^{\Gamma - 1} \mu^n$, where $\Gamma$ is a critical exponent 
which depends on the lattice dimension $D$, and one has $\Gamma = 1$ for $D > 4$
\cite{pr91,so95}. For the SF networks studied here we find 
$s_n \sim \mu_{\infty}^n$, indicating that $\Gamma = 1$, the same exponent as 
for regular lattices in many dimensions.

\subsection{End-to-end separation}
For walks on regular lattices, one usually considers an end-to-end Euclidean 
distance. Our SF networks, however, lack a metric and a true distance is not
defined. Thus, we will consider the end-to-end separation for SAWs on SF
networks as a function of the walk length $n$, the separation between two 
nodes being the number of links along the shortest path connecting them. 
In Fig. 2 we present the average separation $\langle d_n \rangle$ from the
$n$'th node in a SAW to the starting one ($n = 0$) for SF networks
with $\gamma = 3$, $k_0 = 3$, and several sizes $N$. 
This average separation $\langle d_n \rangle$ increases first linearly with $n$
and finally saturates to a finite value that depends on the system size. 
This is logical if one takes into account that these networks are locally 
tree-like, and for the first steps the minimum separation between nodes $n$
and 0 is $d_n = n$. As $n$ increases, there appear shorter ways
connecting nodes $n$ and 0, and finally $\langle d_n \rangle$ becomes
independent of $n$. This saturation for relatively small values of $n$ is
consistent with the small average separation between pairs of nodes (diameter) 
in this kind of networks. In fact, it is known that the diameter increases 
slowly as $\ln N / \ln\ln N$ for $\gamma = 3$, and even more slowly as
$\ln\ln N$ for $\gamma < 3$ \cite{co03}. The limit of $\langle d_n \rangle$ for 
large $n$ in our SAWs is lower than the diameter of the considered networks, 
because nodes with larger $k$ have a higher probability of being visited 
than those with smaller $k$, as indicated above. 
Since the average separation of a node to all other nodes in 
a network decreases for increasing degree of the selected node, the average
separation $\langle d_n \rangle$ for large $n$ in SAWs is lower than the diameter
of the network.
For example, for two of the system sizes represented in Fig. 2  
($N = 2.1 \times 10^3$ and $7.7 \times 10^4$) we have for large $n$,
$\langle d_n \rangle$ = 4.60 and 5.88 vs diameters of 4.77 and 6.07, respectively. 

For regular lattices, the mean squared end-to-end distance of SAWs scales for
large $n$ as $n^{2 \nu}$, $\nu$ being a dimension-dependent critical exponent. 
The upper critical dimension for these walks is $D = 4$,\cite{sl88} which means 
that above this dimension one has $\nu = \frac12$, as for brownian motion 
(Markovian random walks). On the other side, for the SF networks studied here, 
the mean squared end-to-end separation scales as $n^2$ in the thermodynamic limit, 
i.e., with an exponent $\nu = 1$. This exponent coincides with that corresponding 
to $D = 1$, reflecting the fact that loops become irrelevant in the considered
networks as $N \to \infty$ (networks become tree-like). This behavior 
is, however, not captured by SAWs on finite SF networks, for which 
$\langle d_n^2 \rangle$ converges to a constant, and therefore gives a null 
exponent for sufficiently large $n$. 

\begin{figure}
\vspace*{-1.7cm}
\centerline{\psfig{figure=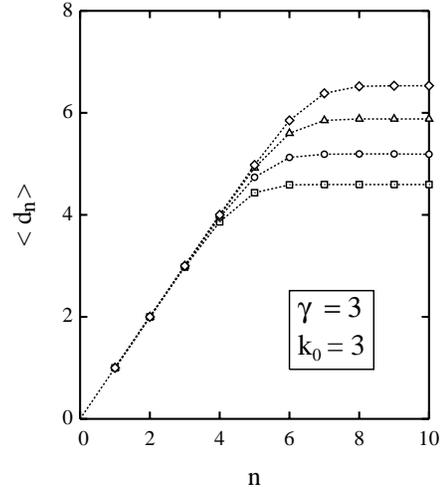,height=10.0cm}}
\vspace*{-1.5cm}
\caption{
Average separation between the $n$'th node and the starting one, for
kinetically-grown SAWs on scale-free networks with $\gamma = 3$ and $k_0 = 3$.
Symbols correspond to different system sizes. From bottom to top:
$N = 2.1 \times 10^3, 9.6 \times 10^3, 7.7 \times 10^4$ and $6.2 \times 10^5$.
Dotted lines are guides to the eye.
} \label{f2} \end{figure}

\section{Self-intersection length}
As indicated above, finite-size effects on SAWs on any (finite) network will
be appreciable as soon as the walks are long enough, as a consequence of the
presence of loops in the network. 
Thus, SAWs are a suitable tool to probe the large-scale topological structure
of complex networks. In particular, the probability of a walk intersecting
itself will depend on the system size, as well as on the topology of the network 
under consideration.
To study this probability, we consider here non-reversal self-avoiding 
walks \cite{so95}, that stop when they try to visit a node already visited 
in the same walk.
The number of steps of a given walk before intersecting itself will be called
`self-intersection length' of the walk, and will be denoted $l$.

To obtain the mean self-intersection length $\langle l \rangle$ of these walks,
we will calculate the probability that a walk stops at step $n$ ($\ll N$). 
Let us consider for the moment nodes with a given degree $k$.
The average number of nodes with degree $k$ visited after $n$ steps is 
\begin{equation}
V_k = n \; Q(k)  \;  ,
\label{vk}
\end{equation}
and the average number of those yet unvisited is
\begin{equation}
U_k = N_k - V_k = N P(k) - n Q(k)  \;  .
\label{uk}
\end{equation}
Then, the probability of reaching in step $n$ an unvisited node is
$u_k \propto k U_k$, and that of finding one already visited is 
$v_k \propto (k-2) V_k$. This is due to the fact that a visited node has 
$k-2$ possible links to reach it, as two of its connections are not available 
because they were employed earlier: one for an incoming step and one for an 
outgoing step.   Therefore, the probability $p_n$ of finding in step $n$ a visited 
node with any degree is
\begin{equation}
p_n = \frac{\sum_k v_k}{\sum_k (v_k + u_k)}  \; .
\label{pn1}
\end{equation}

Inserting into Eq. (\ref{pn1})  expressions (\ref{vk}) and (\ref{uk}) for $V_k$
and $U_k$, and keeping terms linear in $n/N$ one has
\begin{equation}
p_n \approx \frac{n}{\langle k \rangle N}  \sum_k (k-2) Q(k)  \;  ,
\label{pn3}
\end{equation}
and finally
\begin{equation}
p_n \approx  w \frac{n}{N}  \; ,
\label{pn4}
\end{equation}
where 
\begin{equation}
w = \frac{\langle k^2 \rangle - 2 \langle k \rangle}{\langle k \rangle^2} \; .  
\label{ww}
\end{equation}
Note that for the networks considered here $w > 0$. In fact, 
$\langle k^2 \rangle - 2 \langle k \rangle > 0$ is the condition to have a giant
component in a network \cite{mo95}.

\begin{figure}
\vspace*{-1.7cm}
\centerline{\psfig{figure=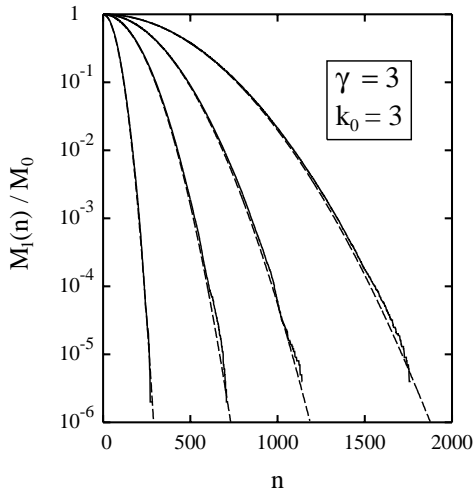,height=10.0cm}}
\vspace*{-1.5cm}
\caption{
Fraction of non-reversal SAWs that survive after $n$ steps,
without intersecting themselves. Results are plotted for SF networks
with $\gamma = 3$, $k_0 = 3$, and several system sizes $N$. From left to
right: $N = 3.3 \times 10^3, 2.6 \times 10^4, 7.7 \times 10^4$, and
$2.1 \times 10^5$. Solid and dashed lines indicate results of numerical
simulations and analytical calculations, respectively.
} \label{f3} \end{figure}
                           
To calculate the probability distribution for the self-intersection length $l$,
we consider $M_0$ random walks starting from nodes taken at random.
We call $M_1(n)$ the number of non-reversal SAWs that remain after $n$ steps
(i.e., those which did not find any node visited earlier).
Thus, $M_1(n) - M_1(n+1) = p_n M_1(n)$, and
considering $n$ as a continuous variable $x$, we have a differential equation
for $M_1(x)$:
\begin{equation}
 \frac{1}{M_1}  \frac{dM_1}{dx}  = - w \frac{x}{N}  \;  ,
\label{diffeq}
\end{equation} 
which yields for integer $n$:
\begin{equation}
 M_1(n) = M_0  \exp  \left( -\frac{w}{2} \frac{n^2}{N} \right) \; .
\label{m1}
\end{equation}

In Fig. 3 we present results for the fraction of surviving walks $M_1(n)/M_0$ 
for SF networks with $\gamma = 3$. We compare the curves derived from 
Eq. (\ref{m1}) (dashed lines) with those obtained from numerical simulations
for networks with $k_0 = 3$ (solid lines). Both sets of results agree with one 
another in the sensitivity region of our numerical procedure (down to $\sim
10^{-6}$). For larger $n$ (and lower $M_1(n)/M_0$) one expects terms of
higher order in $n/N$ to become relevant, and Eq. (\ref{m1}) to be less
reliable.

\begin{figure}
\vspace*{-1.7cm}
\centerline{\psfig{figure=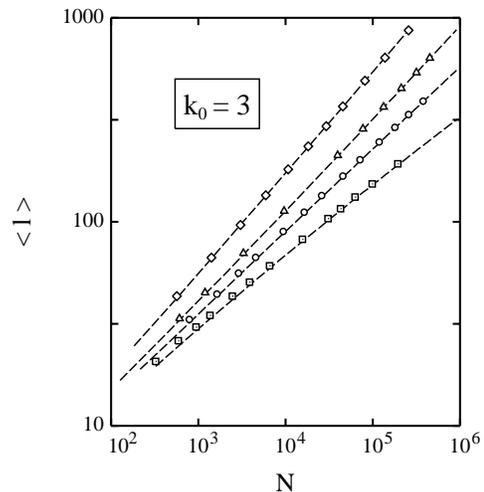,height=10.0cm}}
\vspace*{-1.5cm}
\caption{
Mean self-intersection length $\langle l \rangle$ as a function of system size,
for SF networks with $k_0 = 3$ and several values of the
exponent $\gamma$.  From top to bottom: $\gamma$ = 5, 3, 2.5, and 2.
Symbols are results of numerical simulations and dashed lines were obtained
from Eq. (\ref{avl}).
Error bars of simulation results are less than the symbol size.
} \label{f4} \end{figure}
                           
The average self-intersection length of these walks can be obtained as 
$\langle l \rangle$, with the probability distribution
\begin{equation}
 R(l) =  w \frac{l}{N} \exp \left( -\frac{w}{2} \frac{l^2}{N} \right) \; ,
\label{rl}
\end{equation}
which gives the probability of returning to a visited site in step $l$.
Treating $l$ as a continuous variable, we replace sums by integrals
and find 
\begin{equation}
 \langle l \rangle  \approx \sqrt{ \frac{\pi N}{2 w} }  \; .
\label{avl}
\end{equation}
For large $N$ and $\gamma > 2$, $\langle k \rangle$ converges to a finite
value, and $w \sim \langle k^2 \rangle$; therefore, the mean self-intersection 
length scales as $\langle l \rangle \sim (N / \langle k^2 \rangle)^{1/2}$. 
For $\gamma > 3$, $\langle k^2 \rangle$ does not diverge for large $N$,
and then $\langle l \rangle \sim \sqrt{N}$.
Mean self-intersection lengths derived from Eq. (\ref{avl}) are shown in Fig. 4,
along with those found in numerical simulations for several exponents
$\gamma$.  Both methods give results agreeing with one another within the
error bars of the numerical simulations.
In general, we find $\langle l \rangle \sim N^{\beta}$, with an exponent 
$\beta$ that decreases from 0.5 to 0.25 as $\gamma$ decreases from 3 to 2.
From the distribution for $l$ given by Eq. (\ref{rl}) one finds a mean-square
deviation for the self-intersection length of the walks: $\sigma_l^2 = C N / w$,
with a constant $C = 2 - \pi/2$.  This means that 
$\sigma_l / \langle l \rangle \approx 0.52$. 

We note that the probability distribution $R(l)$ (and therefore the average
value $\langle l \rangle$) is independent of the minimum degree $k_0$.
It depends, apart from the system size, on the exponent $\gamma$ of the 
degree distribution through the mean values $\langle k \rangle$ and
$\langle k^2 \rangle$.
In Fig. 5 we present the mean self-intersection length for SF networks 
with $\gamma = 2.5$ and several values of $k_0$, as derived from numerical
simulations. Different $k_0$ values give a unique $N$-dependence of
$\langle l \rangle$. Thus, the dependence of $\langle l \rangle$ on
$k_0$, which should appear in Eq. (\ref{avl}) through the dependence of
$w$ on $k_0$, is negligible for our purposes.
However, the minimum degree $k_0$ affects strongly other properties of
SAWs, such as the attrition length studied in the following section.

\begin{figure} \vspace*{-1.7cm}
\centerline{\psfig{figure=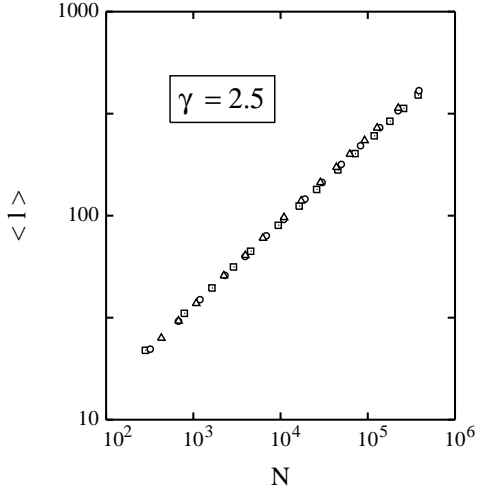,height=10.0cm}}
\vspace*{-1.5cm}
\caption{
Mean self-intersection length $\langle l \rangle$ as a function of system size
for SF networks with $\gamma = 2.5$ and several values of the minimum
degree $k_0$: squares, $k_0 = 3$; circles, $k_0 = 5$; triangles, $k_0 = 7$.
} \label{f5} \end{figure}

\section{Attrition length}
In this section we consider random SAWs, that travel on the network until they 
arrive at a node (called hereafter `blocking node'), where they cannot 
continue because all adjacent nodes have been already visited and are not 
available for the walk. These are kinetic growth walks, as defined by 
Majid {\em et al.} \cite{ma84}.
The number of steps of a given walk until being blocked will be called
attrition length of the walk, and will be denoted $L$.

We now calculate the average attrition length of kinetic growth walks, and 
obtain its asymptotic dependence for large system size $N$. 
To find this average length we will derive a probability distribution for 
$L$, in a way similar to that employed above for the self-intersection length.
With this purpose we note that a blocking node for a kinetic growth walk is 
characterized by
the fact that all its links except one (employed for an incoming step) connect 
it with nodes previously visited.  Then, for step $n$ of a walk and
for a given degree $k$, the average number $N_k'$ of blocking nodes present
in the network is given by the binomial formula:
\begin{equation}
 N_k' =  k \; N_k \; p_n^{k-1} (1-p_n)  \approx  k N_k p_n^{k-1}  \;  ,
\label{nk1}
\end{equation}
where $p_n$ ($\ll 1$ for $n \ll N$) is the average fraction of links joining 
a generic node with nodes visited earlier, as given in Eq. (\ref{pn4}).  
This means that $N_k' \sim k^{1-\gamma} p_n^{k-1}$, and then one has 
$N_k' \ll N_{k_0}'$ for $k > k_0$. Thus, the average number of
links connecting the $n$'th node in a walk with nodes already visited coincides,
within our approximation, with $N_{k_0}'$. (Note that there is one such 
link available for each of those $N_{k_0}'$ nodes.)

The probability of finding a blocking node in step $n$ is given by the
ratio $q_n = \sum_k N_k' / N_{\rm end}$, where 
$N_{\rm end} = \langle k \rangle N$ is the total number of ends of links 
in the network. Then, to order $n/N$, we have 
$q_n \approx N_{k_0}'/\langle k \rangle N$.
This approximation relies
on the fact that $p_n = w n / N \ll 1$, which may be unfulfilled when the 
minimum degree $k_0$ is large and the average attrition length can be on the 
order of the system size $N$ (see below). 

To derive the probability distribution for the attrition length $L$, we
consider $M_0$ kinetic growth walks starting from nodes taken at random.
We call $M_2(n)$ the number of walks that survive after n steps.  Then, the 
number of walks finishing at step $n$ (for which $n$ is a blocking one) is 
$M_2(n) - M_2(n+1) = q_n M_2(n)$.
Considering again $n$ as a continuous variable $x$, one has a differential
equation for $M_2(x)$:
\begin{equation}
 \frac{1}{M_2}  \frac{dM_2}{dx}  = - Y \left( \frac{x}{N} \right)^{k_0-1} \;  ,
\label{diffeq2}
\end{equation}
with the network-dependent constant
\begin{equation}
Y = \frac{N_{k_0}}{N} \frac{k_0}{\langle k \rangle} w^{k_0-1}  \; ,
\label{yy}
\end{equation}
and $w$ given in Eq. (\ref{ww}).  Then, for integer $n$ we have
\begin{equation}
 M_2(n) = M_0  \exp  \left[ - \left( \frac{n}{x_0} \right)^{k_0} \right]   \; ,
\label{m2}
\end{equation}
which gives the number of walks that remain after $n$ steps, i.e. 
$M_2(n)/M_0$ is the probability of surviving to length $n$.  Here $x_0$ is 
a number (dimensionless length) given by $x_0^{k_0} = k_0 N^{k_0 - 1} / Y$.

Therefore, the probability distribution $Z(L)$ for the attrition length of 
these walks is
\begin{equation}
 Z(L) =  \frac{k_0}{L} \left( \frac{L}{x_0} \right)^{k_0}  
         \exp  \left[ - \left( \frac{L}{x_0} \right)^{k_0} \right]   \; .
\label{zl}
\end{equation}
This distribution is strongly dependent on the minimum degree $k_0$, since nodes 
with this degree are in fact controlling the maximum length of kinetic growth
walks in scale-free networks.
The distribution $Z(L)$ is displayed in Fig. 6 for $\gamma = 3$, $k_0 = 3$,
and different system sizes. Dashed lines were obtained from Eq. (\ref{zl}),
and solid lines were derived from numerical simulations. Both sets of results
follow the same trend, but the numerical results seem to be larger than the 
analytical ones for large $L$. This difference is larger than the noise of
the simulation results, and shows the validity limit of our approximation for
large $L$. 

\begin{figure} 
\vspace*{-1.7cm}
\centerline{\psfig{figure=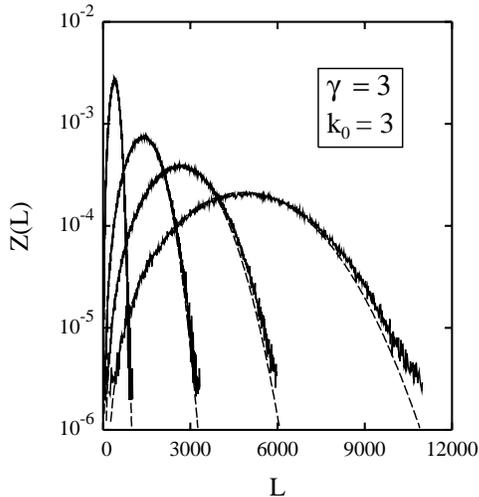,height=10.0cm}}
\vspace*{-1.5cm}
\caption{
Distribution probability $Z(L)$ for the attrition length $L$ of kinetic growth
walks on scale-free networks with $\gamma = 3$, $k_0 =3$, and different system
sizes.
From left to right: $N = 3.3 \times 10^3, 2.6 \times 10^4, 7.7 \times 10^4$,
and $2.1 \times 10^5$.
Solid and dashed lines indicate results of numerical simulations and analytical
calculations, respectively.
} \label{f6} \end{figure}

From the distribution $Z(L)$ we obtain an average attrition length
\begin{equation}
\langle L \rangle \approx \frac{x_0}{k_0} \; \Gamma \left( \frac{1}{k_0} \right)  \; ,
\label{avL}
\end{equation}
$\Gamma$ being Euler's gamma function. Thus, the dependence of $\langle L \rangle$
on $N$ for large systems is controlled by $x_0$. 
To obtain the asymptotic dependence of $x_0$, we note that $Y$ in Eq. (\ref{yy})
scales for $\gamma > 2$ as $Y \sim w^{k_0 - 1}$, because $N_{k_0}/N$ converges to a
constant for large $N$. In addition, $w \sim \langle k^2 \rangle$, and therefore
$x_0^{k_0} \sim  (N / \langle k^2 \rangle)^{k_0 - 1}$.
For $\gamma > 3$, $\langle k^2 \rangle$ converges to a finite value as 
$N \to \infty$, and the average attrition length increases as 
$\langle L \rangle \sim N^{1-1/k_0}$. 
For $\gamma = 3$, $\langle k^2 \rangle \sim \ln N$, and 
$\langle L \rangle \sim (N/\ln N)^{1-1/k_0}$
This means that for a given system size, the average number of nodes visited 
in kinetic growth walks rises with increasing
$k_0$, as a consequence of the increase in the average degree $\langle k \rangle$.  
For $\gamma < 3$, we have $\langle L \rangle \sim N^{\alpha}$, with an
exponent $\alpha$ that decreases from $1-1/k_0$ to $(1-1/k_0)/2$ as 
$\gamma$ decreases from 3 to 2.

In Fig. 7 we show the average attrition length $\langle L \rangle$ as a
function of the system size $N$ for $\gamma = 3$. Symbols correspond to results
of numerical simulations for several values of the minimum degree $k_0$, and 
dashed lines were obtained by using Eq. (\ref{avL}). For the largest $k_0$,
$\langle L \rangle$ derived from simulations increases with $N$ slightly faster 
than the analytical result. This difference is not strange if one observes
that for $k_0 = 9$, $\langle L \rangle$ is on the order of $N$ (in fact, 
for $N = 10^5$, $N / \langle L \rangle \approx 3$), and our assumption 
that $n \ll N$ for all steps of SAWs is not true. However, even in
this case Eq. (\ref{avL}) gives a rather good approximation for the average
length $\langle L \rangle$ (see Fig. 7).

\begin{figure} \vspace*{-1.7cm}
\centerline{\psfig{figure=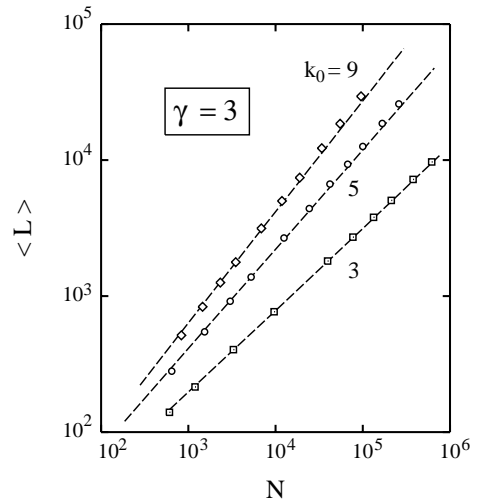,height=10.0cm}}
\vspace*{-1.5cm}
\caption{
Mean attrition length $\langle L \rangle$ as a function of system size for SF
networks with $\gamma = 3$ and several values of $k_0$. From top to bottom:
$k_0$ = 9, 5, and 3. Symbols are results of numerical simulations, and dashed
lines were obtained by using Eq. (\ref{avL}).
Error bars of simulation results are less than the symbol size.
} \label{f7} \end{figure}

In order to define strategies to search in this kind of networks, it is clear 
that nodes with low degree limit the effectiveness of the process. Thus,
actual strategies based on SAWs have to include additional conditions to
improve their efficiency. In this line, Adamic {\em et al.} \cite{ad01} have
proposed an algorithm based on SAWs that prefer high-degree nodes to low-degree 
ones. In any case, the long-range properties of pure SAWs give
us direct insight into the structure of SF networks, further than
the local neighborhood of a node, where the structure of links is tree-like. 
On a larger scale, one always finds loops sooner or later in finite
networks, which is in fact probed by SAWs.  In particular, the average 
self-intersection length $\langle l \rangle$ given in Eq. (\ref{avl}) 
is a measure of the typical size of loops in equilibrium SF networks. 
The presence of loops in a network is responsible for attrition
of the walks. Then, the mean attrition length $\langle L \rangle$ given in
Eq. (\ref{avL}) is a measure of the long-range `openness' of a network
(the longer $\langle L \rangle$, the less loops contains a network). 
In this sense, our SF networks become more `open' as their size $N$ 
increases, and eventually are loop-free (or tree-like) in the thermodynamic 
limit, where both $\langle l \rangle$ and $\langle L \rangle$ diverge to 
infinity.

As a result, we find that the efficiency of SAWs to explore scale-free
networks increases for increasing exponent $\gamma$. 
This is a consequence of the fact that for a given system size $N$, the
fraction of nodes with high degree increases for decreasing $\gamma$.
High-degree nodes are visited more probably than low-degree ones, and
once visited the former are more effective to block a SAW in later steps
(they have more connections), thus reducing the mean self-intersection 
and attrition lengths.

\section{Conclusions}
 Self-avoiding walks provide us with an adequate tool to study the 
long-range characteristics of SF networks. In particular,
they allow us to study the quality of a network to be explored without 
returning to sites already visited.
For large networks, the number of SAWs increases as 
$s_n/s_{n-1} \approx \langle k^2 \rangle / \langle k \rangle - 1$, provided that 
$n \ll N$.
For a given $n$, $s_n$ decreases with decreasing system size,
as a consequence of the presence of loops in the networks.
These finite-size effects affect strongly the maximum length of kinetic
growth walks on scale-free networks.

We have calculated self-intersection and attrition lengths by using an
approximate probabilistic method, which yields results in good agreement
with those derived from numerical simulations.
Both, the average self-intersection length and attrition length scale as a
power of the system size $N$.
For the mean self-intersection length of non-reversal SAWs we have 
$\langle l \rangle \sim N^{\beta}$, with $\beta$ depending on the exponent
$\gamma$ of the degree distribution. In particular, for $\gamma > 3$
one has $\beta = 0.5$, and decreases as $\gamma$ is lowered.
The length of kinetic growth walks in scale-free networks is limited by 
attrition of the paths, and the mean attrition length follows a dependence
$\langle L \rangle \sim N^{\alpha}$, with $\alpha$ depending on $\gamma$
and the minimum degree $k_0$. For $\gamma > 3$, one has
$\alpha = 1 - 1/k_0$.
This dependence of the exponent $\alpha$ on $k_0$ is remarkable, reflecting
the fact that the length of SAWs is limited by attrition at sites with
the minimum degree $k_0$.  \\ 

{\bf Acknowledgments}:
Thanks are due to M. Saboy\'a for critically reading the manuscript. This 
work was supported by CICYT (Spain) under Contract No. BFM2003-03372-C03-03. \\

\end{multicols}

\end{document}